\documentclass[pdftex,12pt,a4paper]{article}
\usepackage{amsmath}  
\usepackage{amssymb}
\usepackage{amsfonts}
\usepackage{mathptmx}
\usepackage{latexsym}
\usepackage{setspace}
\usepackage{graphicx,subfigure}
\usepackage{color}
\usepackage{enumerate}
\usepackage{enumitem}
\usepackage[ocgcolorlinks]{hyperref}
\hypersetup{
  citecolor=blue,
 linkcolor=blue,
 backref=true,pagebackref=true
 }
\usepackage{fancyhdr} 
\begin{document}

\title{\bf Contrasting behavior of the structural and magnetic properties in Mn- and Fe-doped In$_2$O$_3$ films}
\date{}
\maketitle
\vspace{-1in}
{\centering {\bf Qi Feng, Harry J Blythe, A Mark Fox, and Gillian A Gehring\footnote{g.gehring@sheffield.ac.uk}} \\{\it Department of Physics and Astronomy, The University of Sheffield, S3 7RH, UK} \\{\bf Feng-Xian Jiang, and Xiao-Hong Xu}\\{\it School of Chemistry and Materials Science, Shanxi Normal University, Linfen 041004, People's Republic of China}\\{\bf  Steve M Heald}\\{\it Advanced Photon Source, Argonne National Laboratory, Argonne, Illinois, 60439, USA}}

\date{}
{\bf Keyword:{\it Dilute Magnetic Oxide, Magnetism, Magnetic Circular Dichroism, Grain Boundaries}}

\begin{abstract}

We  have observed room temperature ferromagnetism (FM) in In$_2$O$_3$ thin films doped with either 5 at.\% Mn or Fe, prepared  by pulsed laser deposition (PLD)  at substrate temperatures ranging from  300 to 600$\,^{\circ}{\rm C}$. The dependence of   saturation magnetization on   grain size was investigated for both  types of In$_2$O$_3$ films. It is revealed that, for the Mn-doped films,  the magnetization was largest  with small grains,  indicating the importance of grain boundaries. In contrast, for  Fe-doped films,  the largest magnetization was observed with large grains.

\end{abstract}
 
\newpage

Dilute magnetic oxides (DMOs) that exhibit FM  at or above room temperature have attracted considerable attention due to their potential applications in spintronic devices  \cite{Wolf.Sci.2001, Zutic.2004,   Bader.Annu.Rev.Cond.Mat.Phys.2010}. Various transparent oxide thin films have been investigated extensively as host compounds, since these semiconductors can be doped with transition metals (TM) to achieve a high Curie temperature ($T_C$) \cite{Toyosaki.Nat.Mat.2004, Behan.PRL.2008, Fitzgerald.PRB.2006, He.APL.2005, Kobayashi.PRB.2009}.  However, the origin of the observed FM is still not completely understood, with even the necessity for TM doping being called into question.   It has emerged that the most obvious sources for FM could be either localized electrons in donor states or free electrons in the conduction band. Recently, a new model for defect-related FM  has been developed by Coey {\it et al.,} \cite{Coey.NJP.2010} which is due to a spin-split defect-band populated by charge transfer from a proximate charge reservoir. The  observation of  magnetism in granular films and nanoparticles and its absence in bulk materials and epitaxial films was systematized for Mn-doped ZnO samples by Straumal \emph{et al.} \cite{Straumal.PRB.2009,   Straumal.PSS.2011}. They  concluded that the magnetization was localized at the grain boundaries and hence that ferromagnetism was observed provided that the grains were sufficiently small. This letter investigates whether a similar relationship between grain size and magnetism exists in TM-doped In$_2$O$_3$.

In$_2$O$_3$ is a wide band-gap (3.75 eV) conducting transparent semiconductor with cubic bixbyite crystal structure and can possess a high concentration of $n$-type carriers due to  introduced oxygen vacancies. TM-doped In$_2$O$_3$ thin films have attracted great attention since the discovery of high temperature FM in Cr-doped In$_2$O$_3$ thin films by Philip {\it et al.} \cite{Philip.Nat.Mat.2006}.  In recent years, extensive work has been carried out for a better understanding of the origin of the FM. The carrier-mediated exchange  model was suggested by Yoo {\it et al.} \cite{Yoo.APL.2005} when they first reported   high temperature FM in both bulk and thin-film   In$_2$O$_3$ doped with up to 20\% Fe prepared by solid-state reaction and PLD.  Kohiki {\it et al.} \cite{Kohiki.TSF.2006} and Ohno     {\it et al.} \cite{Ohno.JJAP.2006} attributed the observed FM to the formation of Fe clusters or $\beta$-Fe$_2$O$_3$, while the F-centre exchange model was suggested by Xing  {\it et al.}  \cite{Xing.JAP.2009}.   Despite the observation of ferromagnetic behavior in thin films and nanoparticles, some TM-doped In$_2$O$_3$  bulk materials were observed to be paramagnetic \cite{Peleckis.IEEE.2006, Berardan.JMMM.2008}.

In this paper, we investigate the validity of the Straumal \emph{et al.} \cite{Straumal.PRB.2009, Straumal.PSS.2011} picture for In$_2$O$_3$ by studying the correlation between  grain size and magnetism.  The grain size was varied by growing films at different substrate temperatures. We measured the structural, magnetic, transport, optical and magneto-optical properties of In$_2$O$_3$ doped with Mn or Fe. X-ray diffraction (XRD) and optical absorption measurements allow us to estimate the grain size and surface roughness, whereas the structural perfection around the TM site is studied by extended x-ray absorption fine structure (EXAFS). Magnetic properties were characterized by SQUID magnetometry and magnetic circular dichroism (MCD) which confirms the energy of the magnetic states that gives rise to the magnetism.

Polycrystalline thin films of (In$_{0.95}$TM$_{0.05}$)$_2$O$_3$ (TM = Mn, Fe)  were deposited on $c$-cut sapphire substrates at   temperatures ranging from 300 to 600$\,^{\circ}{\rm C}$  in 10$^{-5}$ Torr of oxygen by  PLD with a XeCl laser (308 nm). The laser was used at a pulse repetition rate of 10 Hz, which gave an energy density up to 400 mJ/pulse. The target-to-substrate distance was  40 mm.  Bulk targets for ablation were prepared by a standard solid-state reaction route, as described elsewhere \cite{Hasan.2013}. Phase identification and structural properties of all the films were characterized by XRD  on a Bruker diffractometer in the $\theta$-2$\theta$ mode using Cu $K\alpha$ radiation ($\lambda$ = 1.5406 \AA). The x-ray absorption fine structure (XAFS) measurements were performed at beamline 20-ID-B at the Advanced Photon Source using a microfocused beam incident on the sample at approximately 5$^{\circ}$ glancing angle. The magnetization measurements were performed in a SQUID magnetometer. The thickness of the films was measured with a Dektak profilometer and lay in the range of 50-150 nm. We measured the temperature dependence of the resistance  and Hall effect  in a continuous-flow helium cryostat in the range 5 - 300 K in  fields up to 1 T. Optical absorption spectra and magneto-optic spectra were measured in Faraday geometry by using a photo-elastic modulator from 2.0 eV to 3.5 eV in magnetic fields up to 1.8 T. Since this energy range is below the band-gap of  In$_2$O$_3$, the magneto-optical response probes the magnetic polarization of any gap states.

\begin{figure}[t]
   \centering
      \includegraphics[width=0.8\textwidth]{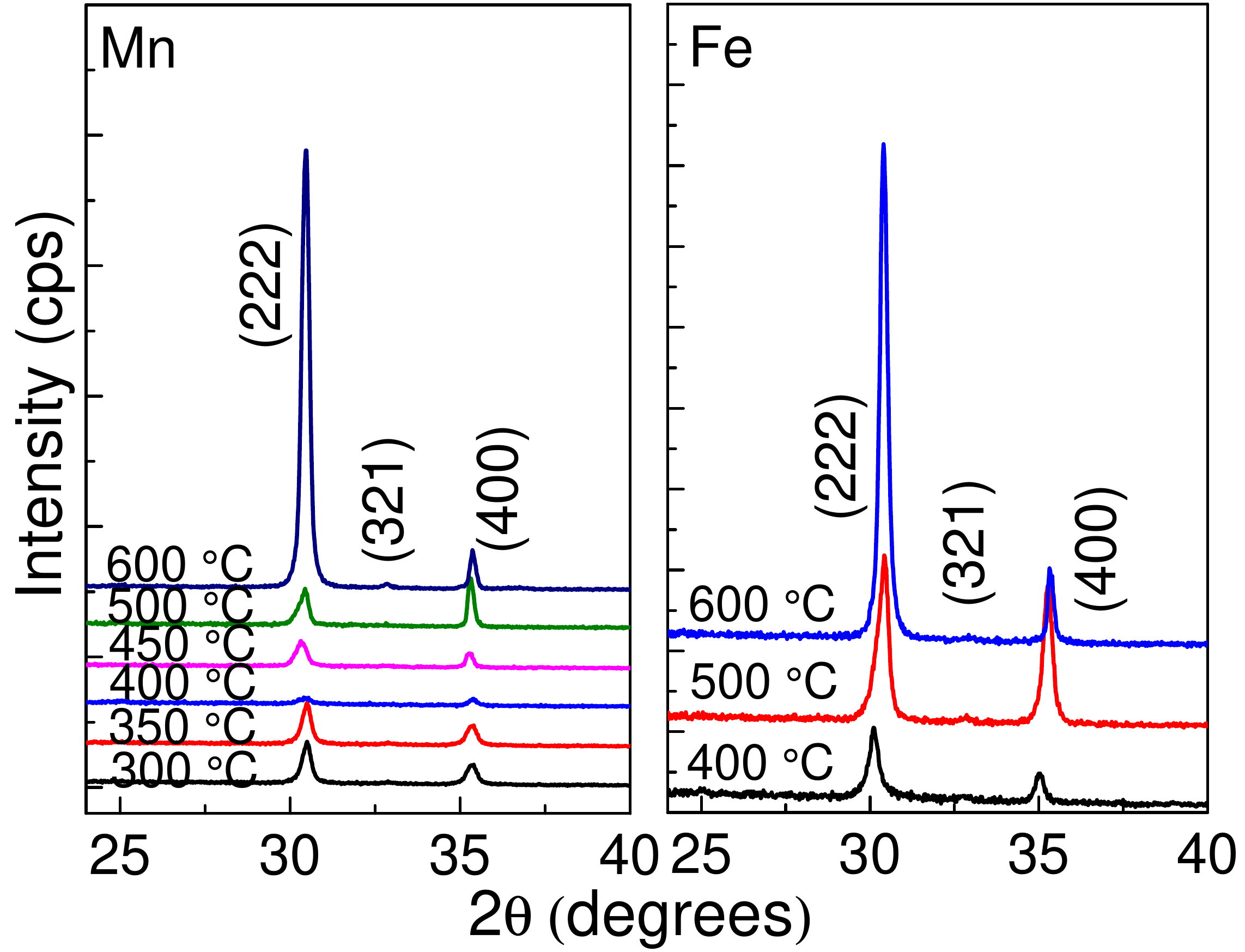} 
       \caption{\label{fig:fig1} (Color Online) $\theta-2\theta$ XRD diffractogram focusing on the In$_2$O$_3$ (222) and (400) reflection peaks for Mn- and Fe-doped In$_2$O$_3$ films grown at different substrate temperatures. Displacement was applied for clarity.  }
\end{figure}

Figure~\ref{fig:fig1} shows the $\theta-2\theta$ XRD scans of Mn- and Fe-doped In$_2$O$_3$ films. All peaks were indexed assuming the same bixbyite cubic structure as  pure In$_2$O$_3$. The lattice constants were comparable to those found for Fe-doped In$_2$O$_3$ \cite{Jiang.JAP.2011}. The average size of the crystalline grains, $D$,  in the films was estimated using the Debye-Scherrer method, $D=\frac{0.94\lambda}{\beta\cos{\theta}}$, where $\lambda$ is the wavelength of the x-ray, $\beta$ the full width at half maximum (FWHM) of the (222) peak and $\theta$ the diffraction angle. The grain size was found to increase with substrate temperature, from 25.2 to 32.8 nm for Mn-doped films, and from 23.5 to 31.1 nm for Fe-doped films, which will be shown later in Fig.\ref{fig:fig5}.

\begin{figure}[t]
   \centering
         \includegraphics[width=0.8\textwidth]{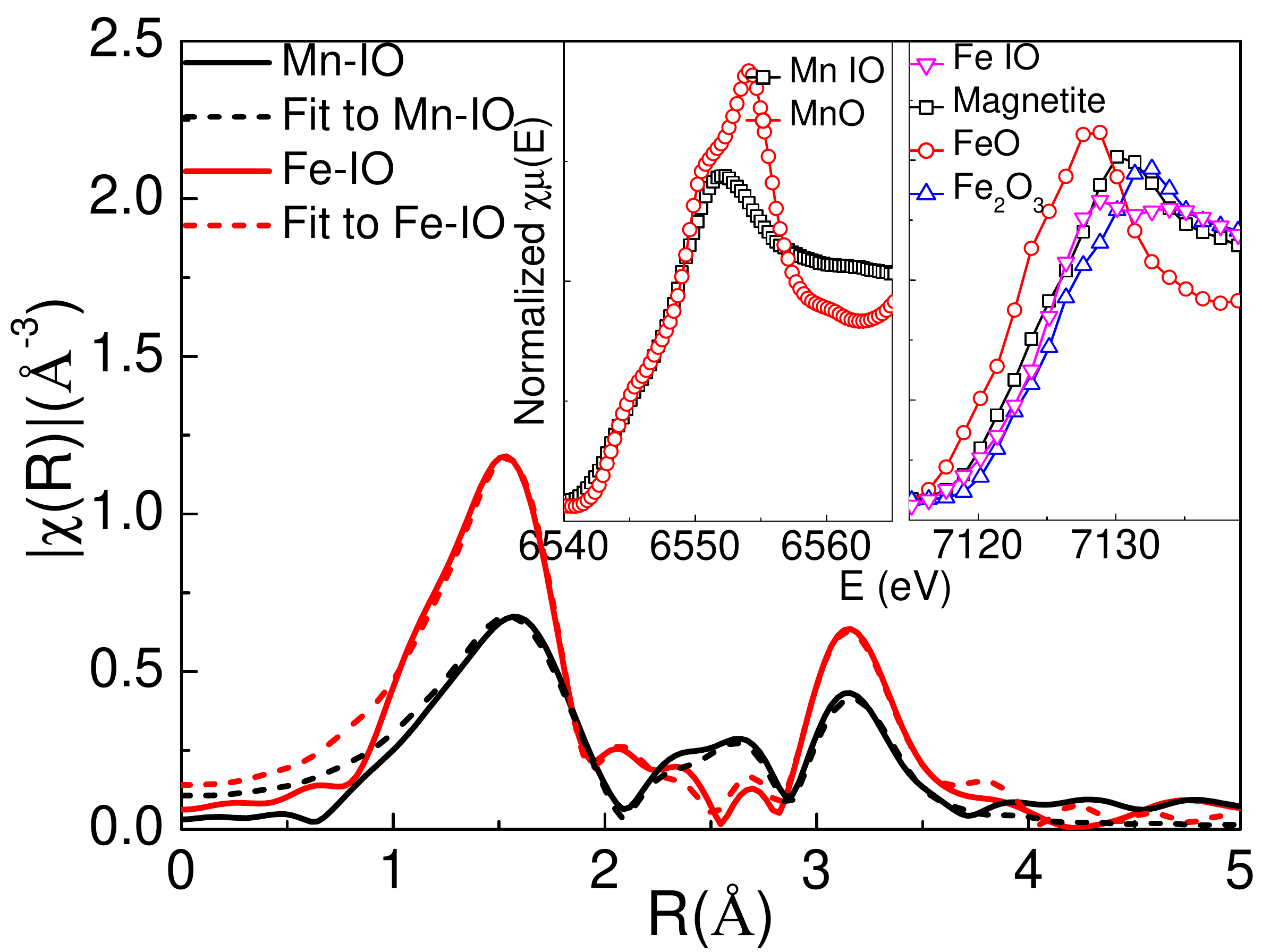} %
       \caption{\label{fig:fig2} (Color Online) $k^2$ weighted Fourier transforms of the EXAFS data (solid lines) along with multi-shell fits to the data (dashed lines) as described in the text. Black and red lines denote Mn-doped In$_2$O$_3$ grown at  400$\,^{\circ}{\rm C}$  and Fe-doped In$_2$O$_3$ sample grown  at 600$\,^{\circ}{\rm C}$, respectively.  The transform range was $k=2-11.5$ \AA$^{-1}$ and the fitting range in $r$ was 1 to 3.7 \AA. A two-site model based in the In$_2$O$_3$ bixbyite structure was used. The Mn fit included 12$\pm$4\% MnO. Insets: Normalized near-edge XAFS spectra for Mn-doped In$_2$O$_3$ which was grown at 400$\,^{\circ}{\rm C}$ and bulk MnO, and Fe-doped In$_2$O$_3$ which was grown at 600$\,^{\circ}{\rm C}$ and different Fe-oxides.}
\end{figure}

XAFS measurements were used to look for potential second phase formation. An analysis was carried out that was very similar to that described earlier \cite{Heald.JPCS.2012}.  As seen in insets in Fig.\ref{fig:fig2} the position of the Mn near edge  indicates clearly that all of the Mn is in the 2+ valence, while the position of the Fe near edge indicates that Fe is in the mixed valence states of Fe$^{2+}$ and Fe$^{3+}$. EXAFS is used to explore the local environment of the doped TM ions. Figure~\ref{fig:fig2}  shows the Fourier transformed data for both the Mn- and Fe-doped samples deposited at 400 and  600 $\,^{\circ}{\rm C}$ respectively along with fits to the data. As shown earlier \cite{Heald.JPCS.2012}, the dopant sites in the In$_2$O$_3$ lattice can have widely varying degrees of disorder.  To fit the data we initially used the  two site model \cite{Heald.JPCS.2012}. This gave a good fit to all the Fe data, suggesting the substitutional sites for Fe, but for the Mn transform extra intensity can be seen between 2-3 \AA\,that was difficult to fit with a pure substitutional model. This is characteristic of a metal oxide impurity, and is due to the strong Mn-Mn signal from typical oxide structures. Therefore, MnO was added to the fitting based on the observed Mn$^{2+}$ valence. A good fit was   obtained for Mn   with the addition of 10-15 \% of second phase MnO for this sample, as shown in Fig.2. For the Mn-doped sample grown at 600 $\,^{\circ}{\rm C}$ the amount of MnO required to fig the data is increased  to 50 \%.

\begin{figure}[t]
    \centering
         \includegraphics[width=0.95\textwidth]{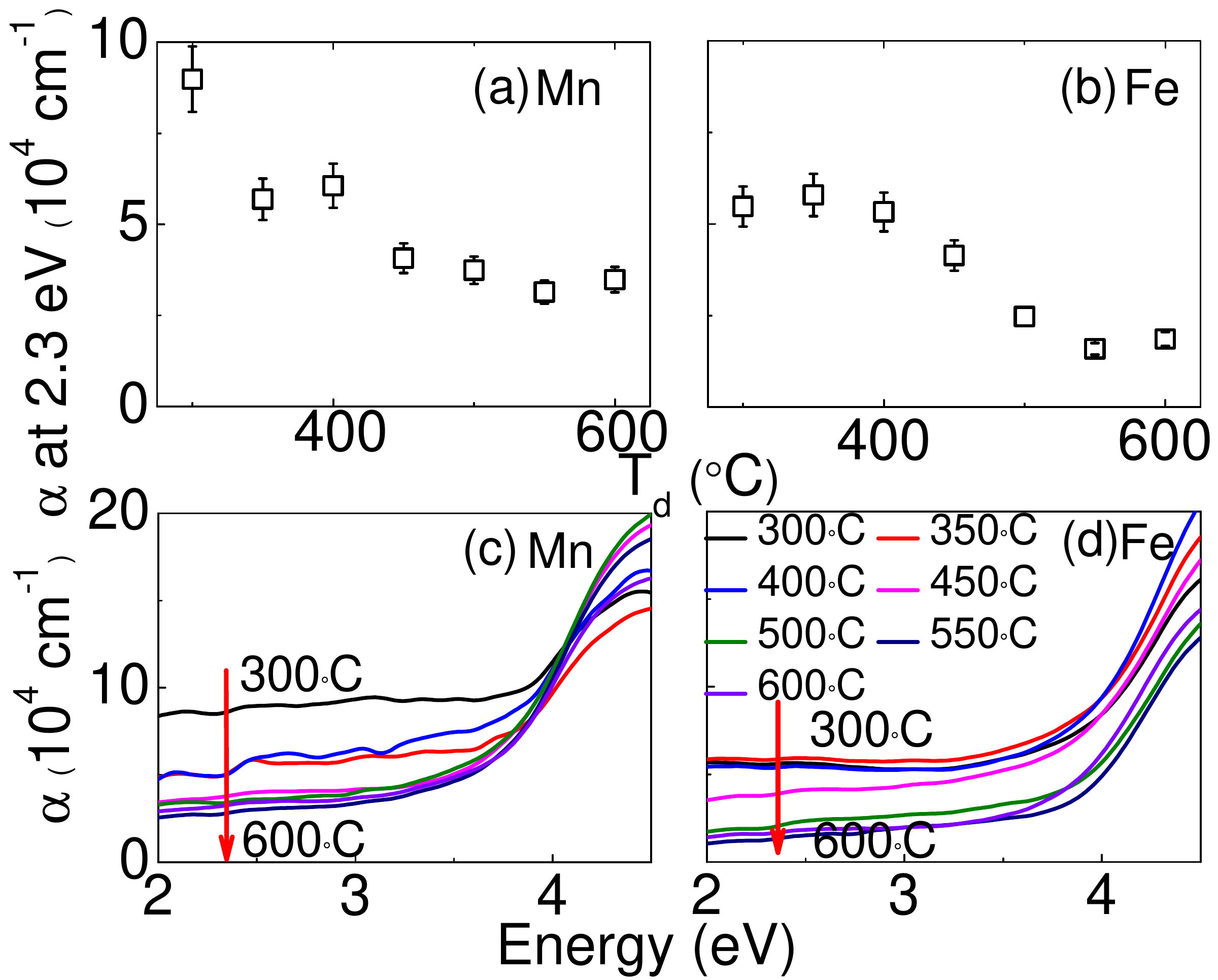}
      \caption{ \label{fig:fig3}(Color Online) (a) and (b) absorption coefficient at 2.3 eV for Mn- and Fe-doped samples, respectively; (c) and (d) room temperature absorption spectra for Mn- and Fe-doped In$_2$O$_3$ films, grown at different substrate temperatures, respectively.}
\end{figure}

Figure~\ref{fig:fig3} shows the optical absorption measurements at room temperature  as a function of photon energy for Mn- and Fe-doped In$_2$O$_3$ films grown at different temperatures.   The absorption in the low energy region decreases monotonically with increasing substrate temperature for both Mn- and Fe-doped samples, as shown in Fig.\ref{fig:fig3}(a) and Fig.\ref{fig:fig3}(b), which is due to the scattering.  The MCD results (shown later) exclude the possibility that it is due to defect phases with smaller band gap and hence these results are a clear indication that the film crystallinity and surface quality increases with increasing deposition temperature.  Thus the optical data is in agreement with the conclusions from the XRD that the grain sizes increase with increasing deposition temperature for both the Mn- and Fe-doped samples.

In summary, the XRD data demonstrates that the grain size increases with the deposition temperature. The EXAFS data shows that all the Fe and most of the Mn is on substitutional sites in In$_2$O$_3$ and that MnO is the only minority phase that is present. Additionally, better crystallinity and surface quality was observed in the optical spectra for films with larger grain size. Below we report on the magnetic properties and relate them to these structural data.

\begin{figure}[t]
\centering
\subfigure{\includegraphics[width=0.7\textwidth]{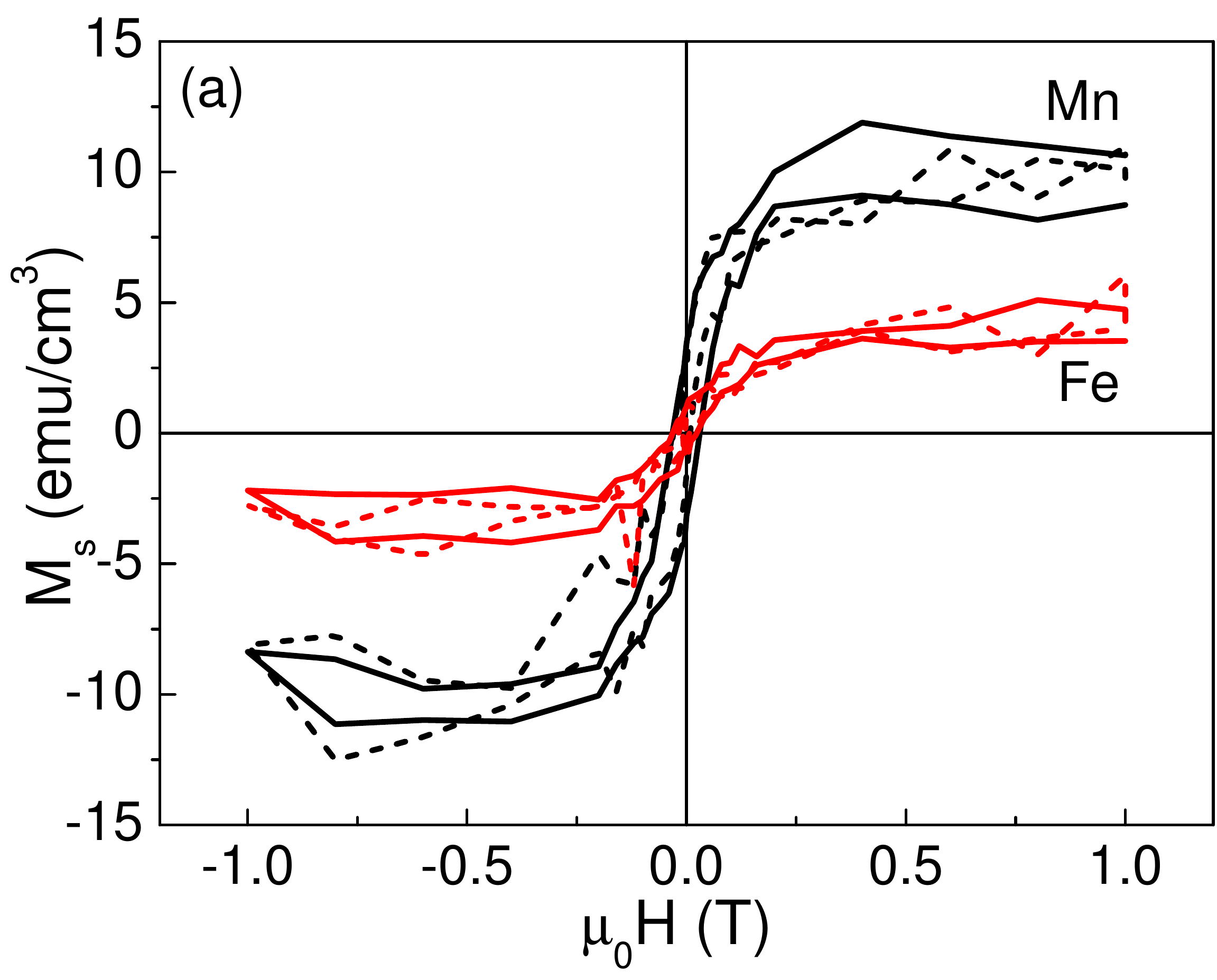}\label{fig:fig4a}}\qquad
\subfigure{\includegraphics[width=0.7\textwidth]{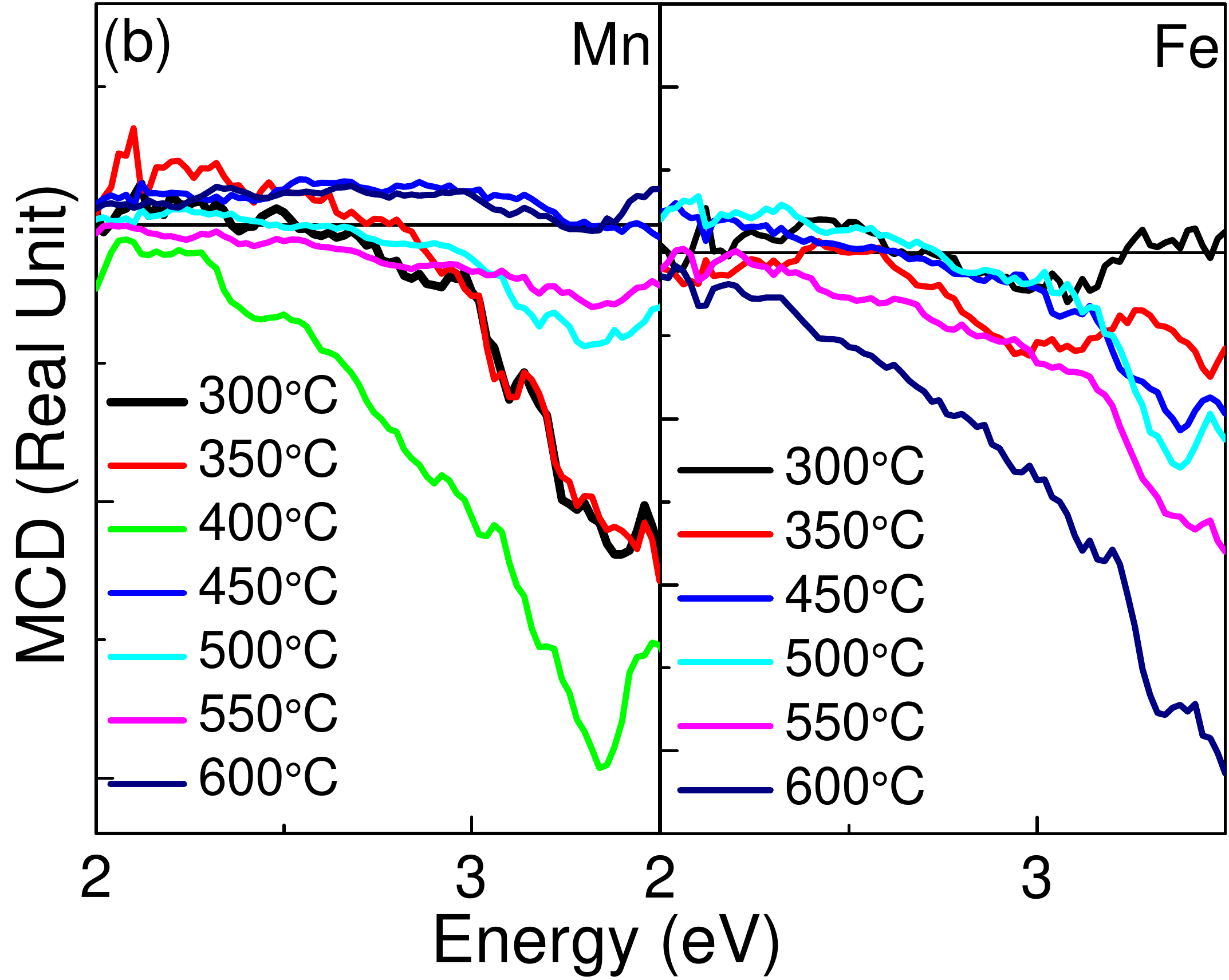}\label{fig:fig4b}}
\caption{ \label{fig:fig4}(Color Online) (a) Room temperature hysteresis loops at 5 and 300 K (dashed and solid lines, respectively) for both Mn- and Fe-doped films (black and red lines, respectively), deposited at 300 and 600 $\,^{\circ}{\rm C}$, respectively, and (b) room temperature MCD curves  as a function of photon energy for Mn- and  Fe-doped In$_2$O$_3$ films deposited  at various temperatures.}
\end{figure}

\begin{figure}[t]
\centering
\subfigure{\includegraphics[width=0.7\textwidth]{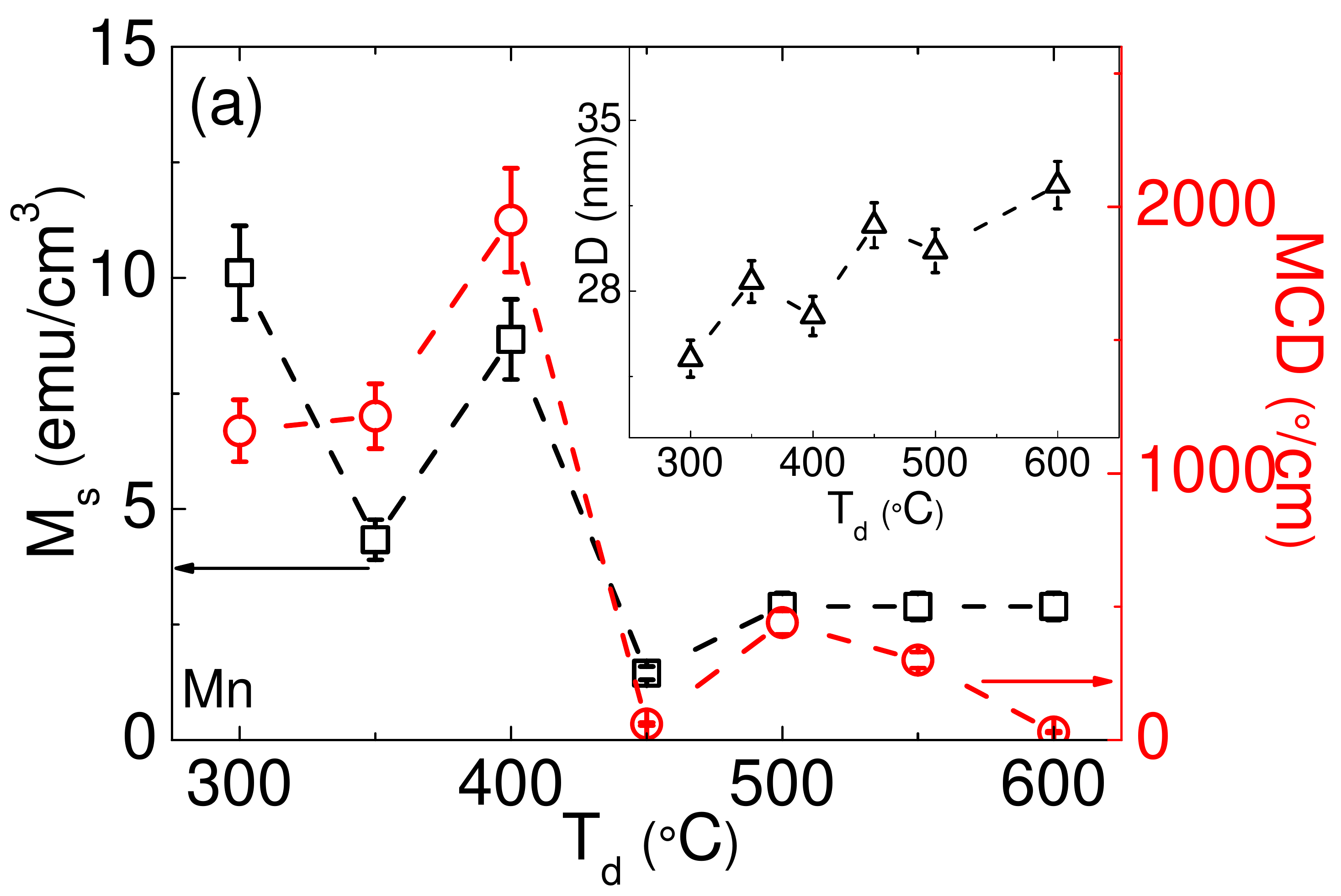}\label{fig:fig5a}}\qquad
\subfigure{\includegraphics[width=0.7\textwidth]{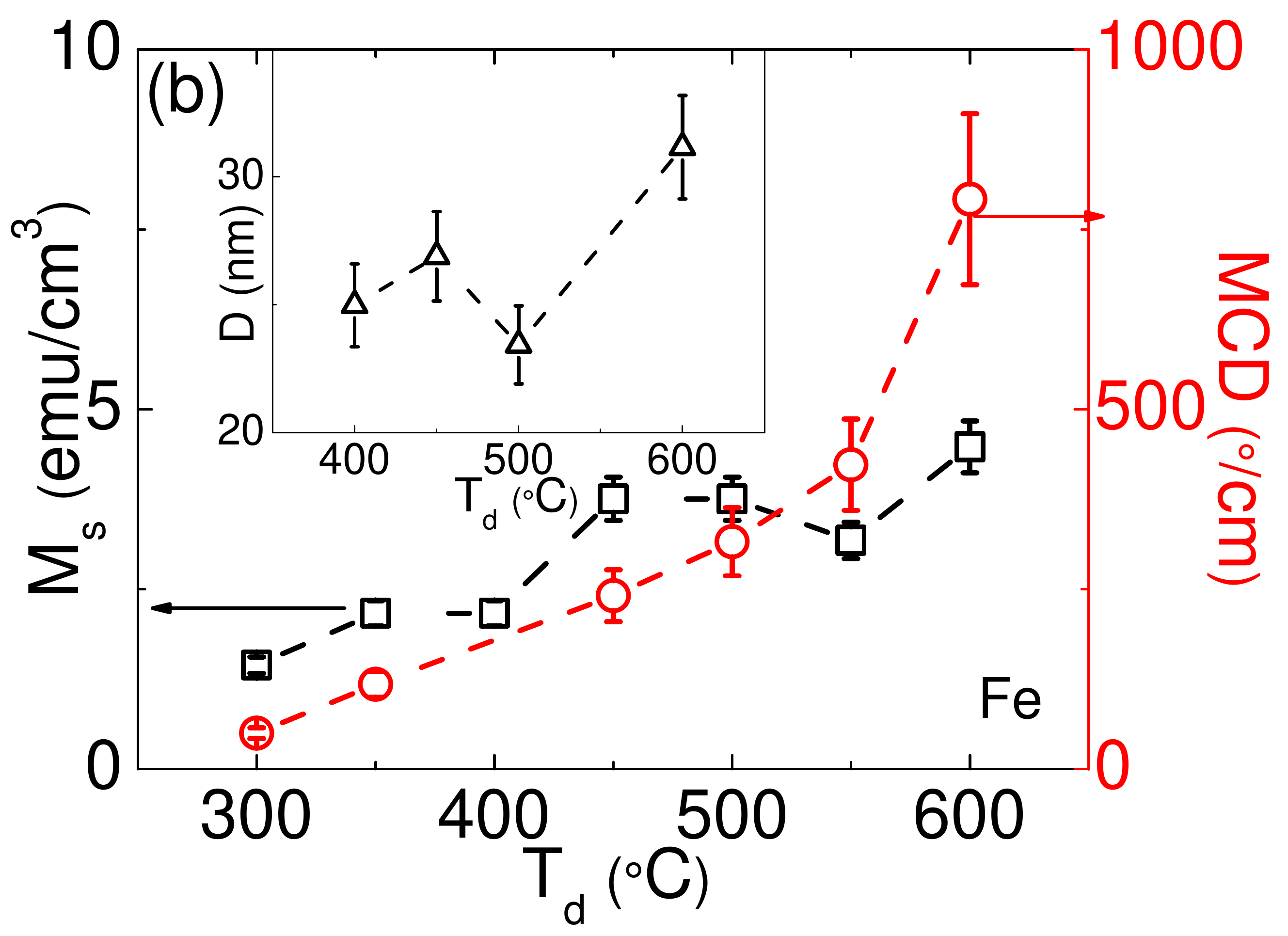}\label{fig:fig5b}}
\caption{\label{fig:fig5}   The variation of   room temperature saturation magnetization, $M_s$ (black squares),  MCD measured at 3.3 eV (red circles), and  grain size, $D$ (triangles) in the insets,  as a function of the substrate temperature ranging from 300 to 600 $\,^{\circ}{\rm C}$ for  In$_2$O$_3$ thin films doped with (5 at.\%) (a) Mn and (b) Fe.}
\end{figure}

Figure~\ref{fig:fig4a} shows the in-plane M-H curves of Mn- and Fe-doped In$_2$O$_3$ films at 5 and 300 K. Similar temperature independent hysteresis loops were observed in all the Mn- and Fe-doped samples. A  strong correlation between the magnetic behavior and transport properties was reported by Jiang {\it et al.} \cite{Jiang.JAP.2011} and Park {\it et al.} \cite{Park.APL.2012} They found that for Fe-doped In$_2$O$_3$ thin films, magnetization was temperature independent in a metallic sample, whereas  temperature dependent magnetization was observed in a semiconducting sample. All of our samples  show metallic behavior with high carrier concentration in the range between 1.00$\times$10$^{20}$ and 7.45$\times$10$^{20}$ cm$^{-3}$ and $M_s$ is independent of  temperature, in agreement with Ref \cite{Jiang.JAP.2011, Park.APL.2012}.  Figure~\ref{fig:fig4b} shows the room temperature MCD as a function of photon energy for Mn- and Fe-doped In$_2$O$_3$ films.  The MCD spectra, like the saturation magnetization, were unchanged down to 10 K. The MCD signal is characteristic of a magnetic oxide and the lack of a change in the spectra for the Mn-doped sample below 100 K indicates that the MnO is not contributing to the overall magnetization.

Figure~\ref{fig:fig5a} shows the variation of the room temperature saturation magnetization ($M_s$)  for Mn-doped In$_2$O$_3$ films,  as a function of substrate temperature.  A  large reduction in $M_s$ with substrate temperature was observed. This correlates with  EXAFS data indicating that the fraction of the sample that is MnO rises with temperature. This is the strongest indication that MnO is not responsible for the magnetization observed for films deposited at 300 $\,^{\circ}{\rm C}$ as the magnetization falls dramatically as the fraction of MnO is increased.  The possibility of Mn double exchange interactions may be discounted because there is no sign that the Mn is mixed valent. Given that MnO is not contributing directly to the ferromagnetism we note that an increase in MnO corresponds to a drop in the concentration of Mn$^{2+}$ in In$_2$O$_3$. We conclude that the observed decrease in magnetization was too rapid to be accounted for by this mechanism.

The  MCD spectra   at 3.3 eV is also shown in Fig.\ref{fig:fig5a}, which is just below the band-gap; this is believed to correspond to transitions to spin-polarized gap states due to   oxygen vacancies or grain boundaries. The MCD as a function of substrate temperature behaves similarly  to that of the saturation magnetization obtained from SQUID magnetometry. It is strong evidence to support the contention that the observed FM  originates from the same spin-polarized states as those involved in the optical transitions.

Figure~\ref{fig:fig5b} shows the corresponding data for Fe-doped films. Different behavior of both the MCD and $M_s$ as a function of substrate temperature    was observed. In Fe-doped films, Fe is in the substitutional sites as mixed valence states of Fe$^{2+}$ and Fe$^{3+}$. Both the grain size and magnetization were found to increase with deposition temperature. In contrast, in Mn-doped films, the grain size also increases as the deposition temperature increases but the magnetization decreases. The significant differences  between Mn- and Fe-doped films suggested that,   for Mn-doped films, the MCD is due to the electronic states which are localized at the grain boundaries,  whereas for Fe-doped films  the MCD is due to the electronic states in the grains.

In conclusion, we have observed room temperature ferromagnetism  in both  Mn- or  Fe-doped In$_2$O$_3$ films by SQUID and MCD measurements. A  similar substrate temperature dependence of the grain size was observed in  Mn- and Fe-doped In$_2$O$_3$ films which correlated with contrasting magnetic properties. The large magnetization seen for Mn-doped In$_2$O$_3$ when the grains were small indicates that grain boundary FM is responsible for the FM  in Mn-doped  In$_2$O$_3$ films, similar  to that found in Mn-doped ZnO films \cite{Straumal.PRB.2009}. In contrast, we found that the magnetism in Fe-doped films was increased when the fraction of boundaries was reduced and originates from Fe ions in the grains.   The present work  indicates that Fe-doped In$_2$O$_3$ films behave intrinsically and they are more suitable than Mn-doped In$_2$O$_3$ thin films for applications.

Use of the Advanced Photon Source, an Office of Science User Facility operated for the U.S. Department of Energy (DOE) Office of Science by Argonne National Laboratory, was also supported by the U.S. DOE under Contract No. DE-AC02-06CH11357. We thank Dr. Xiufang Qin for help with the XRD measurements.


\newpage

\clearpage

\clearpage

\clearpage

\clearpage

\end{document}